\documentclass[aps,prl,twocolumn,groupedaddress,amssymb,showpacs,epsfig]{revtex4}
\usepackage{graphicx}
\usepackage{bm}
\usepackage{dcolumn}
\usepackage{amsmath}
\usepackage{amssymb}
\usepackage{epsfig}

\def \lleq {\lower0.9ex\hbox{ $\buildrel < \over \sim$} ~}
\def \ggeq {\lower0.9ex\hbox{ $\buildrel > \over \sim$} ~}

\def \omx   {\Omega_{DE}}
\def \om   {\Omega_m}

\def \omm  {\Omega_{0 {\rm m}}}

\def \beq  {\begin{equation}}
\def \eeq  {\end{equation}}
\def \ber  {\begin{eqnarray}}
\def \eer  {\end{eqnarray}}

\def\apj{{Astroph.\@ J.\ }}
\def\mn{{Mon.\@ Not.\@ Roy.\@ Ast.\@ Soc.\ }}

\def\pd{{Phys.\@ Rev.\@ D\ }}

\def\plb {{Phys.\@ Lett.\@ B\ }}
\def \jetpl {JETP Lett.\ }
\def \cqg {{Class. Quant. Grav.\ }}

\begin{document}
\newcommand{\newc}{\newcommand}

\newc{\be}{\begin{equation}}
\newc{\ee}{\end{equation}}
\newc{\ba}{\begin{eqnarray}}
\newc{\ea}{\end{eqnarray}}
\newc{\bea}{\begin{eqnarray*}}
\newc{\eea}{\end{eqnarray*}}
\newc{\D}{\partial}
\newc{\ie}{{\it i.e.} }
\newc{\eg}{{\it e.g.} }
\newc{\etc}{{\it etc.} }
\newc{\etal}{{\it et al.}}
\newc{\lcdm }{$\Lambda$CDM }
\newcommand{\nn}{\nonumber}
\newc{\ra}{\rightarrow}
\newc{\lra}{\leftrightarrow}
\newc{\lsim}{\buildrel{<}\over{\sim}}
\newc{\gsim}{\buildrel{>}\over{\sim}}

\title{The Statefinder hierarchy:
An extended null diagnostic for
concordance cosmology}

\author{Maryam Arabsalmani$^a$ and Varun Sahni$^a$}
\affiliation{
$^a$ Inter University Centre for Astronomy and Astrophysics, Post Bag 4,
Ganeshkhind, Pune, 411007, India}
\date{\today}

\begin{abstract}
We show how higher derivatives of the expansion factor can be developed into
a null diagnostic for concordance cosmology ($\Lambda$CDM).
It is well known that the Statefinder -- the third derivative of the
expansion factor written in dimensionless form, $a^{(3)}/aH^3$, equals
unity for $\Lambda$CDM. We generalize this result and demonstrate that the
hierarchy, $a^{(n)}/aH^n$, 
can be converted to a form that stays pegged at unity in concordance cosmology.
This remarkable property of the Statefinder hierarchy enables it to be used as 
an extended null diagnostic for the cosmological constant. 
The Statefinder hierarchy combined with the growth
rate of matter perturbations defines a {\em composite null 
diagnostic} which can distinguish evolving dark energy 
from \lcdm.
\end{abstract}
\pacs{98.80.Es,98.65.Dx,98.62.Sb}
\maketitle

\section{Introduction}

Despite its radical connotations, there is mounting observational evidence
in support of a universe that is currently accelerating \cite{observations}. 
Theoretically
there appear to be
two distinct ways in which the universe 
can be made to accelerate \cite{DE_review}: (i) through the presence of an additional component
in the matter sector, which, following \cite{ss06},
 we call {\em physical dark energy}.
Physical DE models possess large negative pressure and lead to
the violation of the strong energy condition, $\rho+3P \geq 0$, which forms a 
necessary condition for achieving cosmic acceleration. 
Prominent examples of this class of models include the cosmological constant 
`$\Lambda$',
Quintessence,
the Chaplygin gas, etc.
(ii) The universe can also accelerate because of changes in the gravitational
sector of the theory. These models (sometimes referred to as {\em geometrical DE} or
modified gravity) include $f(R)$ theories, extra-dimensional Braneworld
models, etc.

Due to its elegance and simplicity the cosmological constant,
with $P = -\rho$, occupies a 
privileged place in the burgeoning pantheon of DE models.
Although the reasons behind the extremely small value of $\Lambda$ remain unclear,
concordance cosmology ($\Lambda$CDM) 
does appear to provide a very good fit to current data
(although possible departures from $P = -\rho$ have also been noted \cite{sss09}).
Given the success and simplicity of concordance
cosmology it is perhaps natural to discuss diagnostic measures which
can be used to compare a given DE model with $\Lambda$CDM. `Null measures' of
 concordance cosmology proposed so far include the $Om$ diagnostic 
\cite{Om,zunkel_clarkson}, and the Statefinders \cite{statefinder}.
While $Om$ involves measurements of the expansion rate, $H(z)$, the Statefinders
are related to the third derivative of the expansion factor.
In a spatially flat $\Lambda$CDM universe, 
the Statefinders and $Om$ remain pegged at a fixed
value during our recent expansion history $(z \lleq 10^3)$. 

In this letter we introduce the notion of the `{\em Statefinder hierarchy}' 
which includes higher derivatives of the expansion factor
$d^na/dt^n, n \geq 2$. 
We demonstrate that,
for concordance cosmology, all members of the 
Statefinder hierarchy 
can be expressed in terms of elementary functions of
the deceleration parameter $q$ (equivalently the density parameter 
$\om$). This property singles out the cosmological
constant from evolving DE models and allows the Statefinder hierarchy to be
used as an extended null diagnostic for $\Lambda$CDM.

\section{The Statefinder Hierarchy}
\label{sec1}

The expansion factor of the universe can be Taylor expanded 
around the present epoch $t_0$ as follows
\beq
(1+z)^{-1} := \frac{a(t)}{a_0} = 1 + \sum_{n=1}^\infty\frac{A_n(t_0)}{n!}
\left\lbrack H_0(t-t_0)\right\rbrack^n
\label{scale}
\eeq
where
\beq
A_n:=\frac{a^{(n)}}{aH^n}, ~~~~~n\in N,
\label{statefinder}
\eeq
$a^{(n)}$ is the $n^{th}$ derivative of
the scale factor with respect to time.
Historically different letters of the alphabet have been used to describe
various derivatives of the scale factor. Thus 
$q \equiv -A_2$ is the deceleration parameter, while $A_3$,
which was first discussed in \cite{chiba}, has been called 
the Statefinder `$r$' \cite{statefinder} as well as
the jerk `$j$' \cite{visser}, $A_4$ is the snap `$s$', $A_5$ is the lerk `$l$', etc.
(See for instance \cite{visser,capo,gibbons} and references therein.)

\begin{figure}
\begin{center}
\vspace{-0.1cm}
\psfig{figure=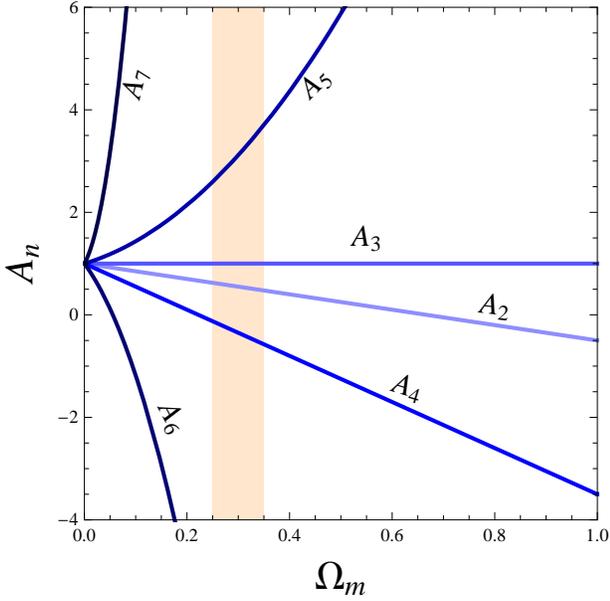,width=0.45\textwidth,angle=0}
\vspace{-0.6cm}
\end{center}
\caption{\small $A_n$ are plotted against $\om(z)$ for \lcdm.
} \label{fig1}
\end{figure}

It is quite remarkable that, in a spatially flat universe consisting of 
pressureless matter and a cosmological constant (henceforth
referred to as concordance cosmology or $\Lambda$CDM),
all the $A_n$ parameters
can be expressed as elementary functions of the deceleration parameter $q$,
or the density parameter $\om$.
For instance $^1$
\footnotetext[1]{It is easy to see that successive differentiations of ${\ddot a} = a({\dot H} + H^2)$
and $R = 6({\dot H} + 2H^2)$ in conjunction with (\ref{state1}),
can be used to express higher derivatives of the Hubble parameter, $H^{(n)}/H^{n+1}$ ,
and the Ricci scalar, $R^{(n)}/6H^{n+2}$, as polynomial expansions in $q$
(or $\om$),
for concordance cosmology.}

\begin{eqnarray}
&&A_2 = 1 - \frac{3}{2}\om,\cr\cr
&&A_3=1,\cr\cr
&&A_4=1 - \frac{3^2}{2}\om,\cr\cr
&&A_5=1+3\om+\frac{3^3}{2}\om^2,\cr\cr
&&A_6=1-\frac{3^3}{2}\om-{3^4}\om^2-\frac{3^4}{4}\om^3, ~~{\rm etc,}
\label{state1}
\end{eqnarray}
where $\Omega_m = \omm (1+z)^3/h^2(z)$ and 
$\Omega_m = \frac{2}{3}(1+q)$ in concordance cosmology.
It is interesting to note that while $A_3$ remains pegged at unity,
the
remaining $A_n$ parameters evolve with time with 
even $A_{2n}$ (odd $A_{2n+1}$) remaining
smaller (larger) than unity, see figure 1.
All $A_n$ approach unity in the distant future:
$A_n \to 1$ when $\om \to 0$ and $\Omega_{\Lambda} \to 1$.
The above expressions allow us to define 
the {\em Statefinder hierarchy} $S_n$:
\begin{eqnarray}
&&S_2 := A_2 + \frac{3}{2}\om,\cr\cr
&&S_3  := A_3,\cr\cr
&&S_4 := A_4+ \frac{3^2}{2}\om,\cr\cr
&&S_5 := A_5-3\om-\frac{3^3}{2}\om^2,\cr\cr
&&S_6 := A_6+\frac{3^3}{2}\om+{3^4}\om^2+\frac{3^4}{4}\om^3, ~~{\rm etc.}
\label{state0}
\end{eqnarray}
The Statefinder 
stays pegged at unity for \lcdm
\beq
S_n\bigg\vert_{\Lambda{\rm CDM}} = 1, 
\label{statefinder1}
\eeq
during the entire course of cosmic expansion !

Equations (\ref{statefinder1}) 
define a {\it null diagnostic} for concordance cosmology,
since some of these equalities are likely to be violated by evolving DE models
for which one might expect one (or more)
of the Statefinders to depend upon time, see figure 2.

\begin{figure*}[!t]
\centering
\begin{center}
\vspace{-0.05in}
\centerline{\mbox{\hspace{0.in} \hspace{2.9in}  \hspace{2.9in} }}
$\begin{array}{@{\hspace{-0.3in}}c@{\hspace{0.3in}}c@{\hspace{0.3in}}c}
\multicolumn{1}{l}{\mbox{}} &
\multicolumn{1}{l}{\mbox{}} \\ [-0.5cm]
\includegraphics[scale=1.0, angle=0]{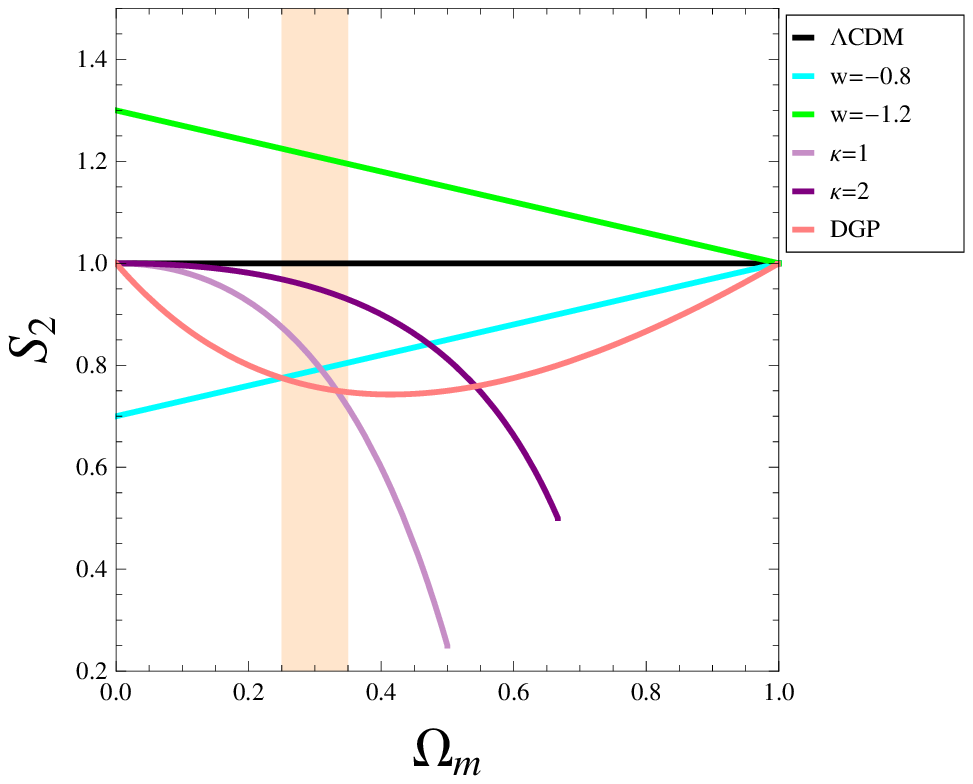}
\includegraphics[scale=1.0, angle=0]{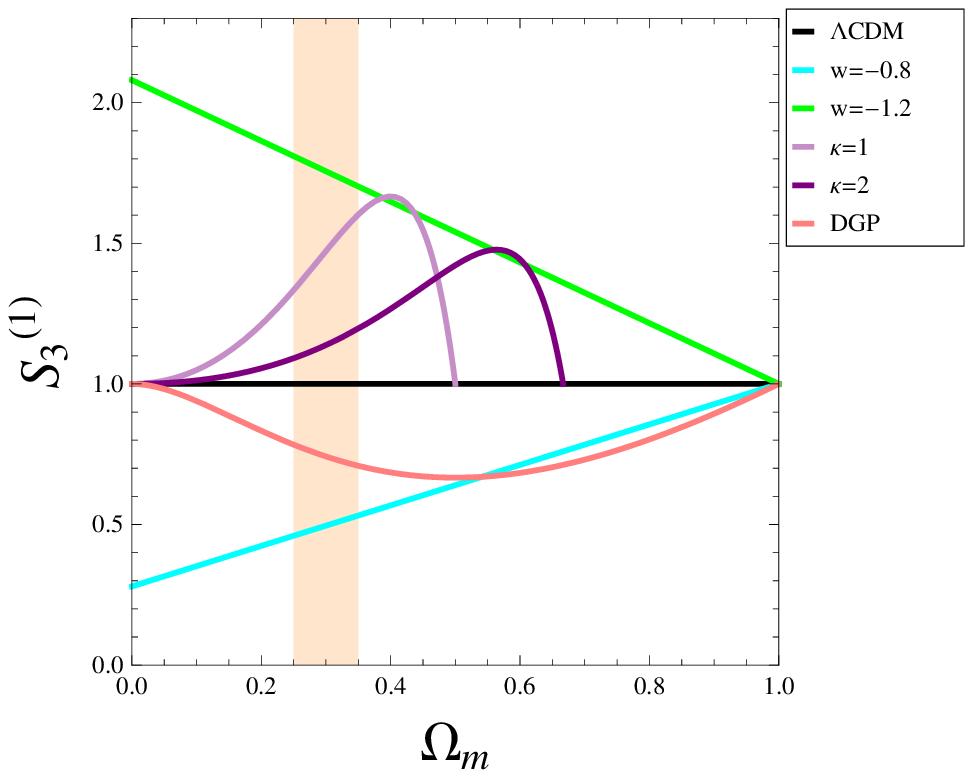}
\end{array}$
\end{center}
\caption{\small
The left (right) panel shows the Statefinder $S_2$
($S_3^{(1)} \equiv S_3$) plotted
against $\om \equiv \omm(1+z)^3/h^2$. Large values $\om \to 1$
correspond to the distant past ($z \gg 1$), while small values 
$\om \to 0$ correspond to the remote future ($z \to -1$).
The models are: DE with $w = -0.8$ (blue), phantom with 
$w = -1.2$ (green),
Chaplygin gas (purple), DGP (red). 
The horizontal black line shows $\Lambda$CDM.
The vertical band centered at $\omm = 0.3$ roughly corresponds to the present 
epoch.
Note that the near degeneracy seen in $S_2$ between 
DGP, $w= -0.8$ and $\kappa=1$ Chaplygin gas,
at $\omm \simeq 0.3$, is absent in $S_3^{(1)}$.
} \label{fig2}
\end{figure*}

\subsection{Fractional Statefinders}
\label{sec2}

Interestingly,
for $n>3$ there is more than one way in which to define a null
diagnostic. 
Using the relationship $\Omega_m = \frac{2}{3}(1+q)$, valid in \lcdm,
it is easy to see that the Statefinders can also be written in the
alternate form 
\ber
&&S_4^{(1)} := A_4 + 3(1+q)\nonumber\\
&&S_5^{(1)} := A_5 - 2(4+3q)(1+q)~, ~~{\rm etc}.
\eer
Clearly both methods yield identical results for  \lcdm: 
$S_4^{(1)} := S_4 = 1$, $S_5^{(1)} := S_5 = 1$.
For other DE models however, the two alternate definitions of the statefinder,
$S_n$ \& $S_n^{(1)}$,
are expected to give different results.

In \cite{statefinder} it was shown that
a second Statefinder could be constructed from the Statefinder
$S_3^{(1)} := S_3$, namely
\beq
S_3^{(2)} = \frac{S_3^{(1)}-1}{3(q-1/2)}~.
\label{eq:second_state}
\eeq
In concordance cosmology $S_3^{(1)} = 1$ while
$S_3^{(2)}$
stays pegged at zero.
Consequently the Statefinder pair $\lbrace S_3^{(1)},S_3^{(2)}\rbrace
 = \lbrace 1, 0\rbrace$ provides a model
independent means of distinguishing evolving dark energy models
from the cosmological constant \cite{statefinder}.
In analogy with (\ref{eq:second_state})
we define the second member of the Statefinder hierarchy as follows$^2$\footnotetext[2]{$\lbrace S_3^{(1)},S_3^{(2)}\rbrace \equiv \lbrace r, s \rbrace$
in \cite{statefinder}. 
Eq. (\ref{state1}) can be used to define still other null tests including:
$Om_1 = \frac{1-\Omega_{0m}}{h^2}+\om(z)$, 
$\Sigma_1 := 2(1-A_4)/9\om$, $\Sigma_2 := A_4 + 3(A_3-A_2)$,
where $Om_1 = 1,$ $\Sigma_{1,2} = 1$ for \lcdm.}
\beq
S_n^{(2)} = \frac{S_n^{(1)}-1}{\alpha(q-1/2)},
\label{state2}
\eeq
where $\alpha$ is an arbitrary constant.
In concordance cosmology $S_n^{(2)} = 0$
and
\beq
\lbrace S_n^{(1)},S_n^{(2)}\rbrace
 = \lbrace 1, 0\rbrace~.
\eeq
The second statefinder $S_n^{(2)}$ serves the useful purpose of
breaking some of the degeneracies present in 
$S_n^{(1)}$.
For DE with a constant equation of state $w$
\ber\label{eq:statefinder1}
S_3^{(1)} &=& 1 + \frac{9w}{2}(w+1) \omx\nonumber\\ 
S_3^{(2)} &=& w+1\nonumber\\
S_4^{(1)}&=& 1-\frac{27}{2}w(w+1)(w+\frac{7}{6})\omx-
\frac{27}{4}w^2(w+1)\omx^2\nonumber\\
S_4^{(2)}&=& -(w+1)(w+\frac{7}{6})-\frac{1}{2}w(w+1)\omx\nonumber
\eer
where $S_4^{(2)}=\frac{S_4^{(1)}-1}{9(q-\frac{1}{2})}$ and
$q-1/2 = \frac{3w}{2}\omx$.

\begin{figure*}[!t]
\centering
\begin{center}
\vspace{-0.05in}
\centerline{\mbox{\hspace{0.in} \hspace{2.0in}  \hspace{2.0in} }}
$\begin{array}{@{\hspace{-0.3in}}c@{\hspace{0.3in}}c@{\hspace{0.3in}}c}
\multicolumn{1}{l}{\mbox{}} &
\multicolumn{1}{l}{\mbox{}} \\ [-0.5cm]
\includegraphics[scale=1.0, angle=0]{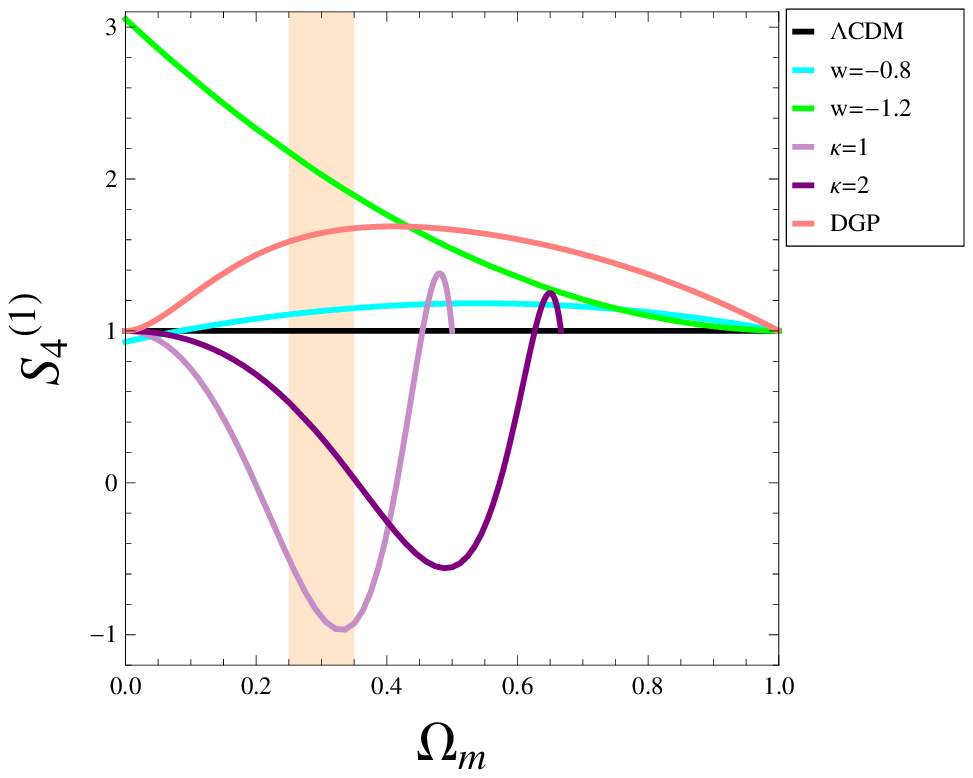}
\includegraphics[scale=1.0, angle=0]{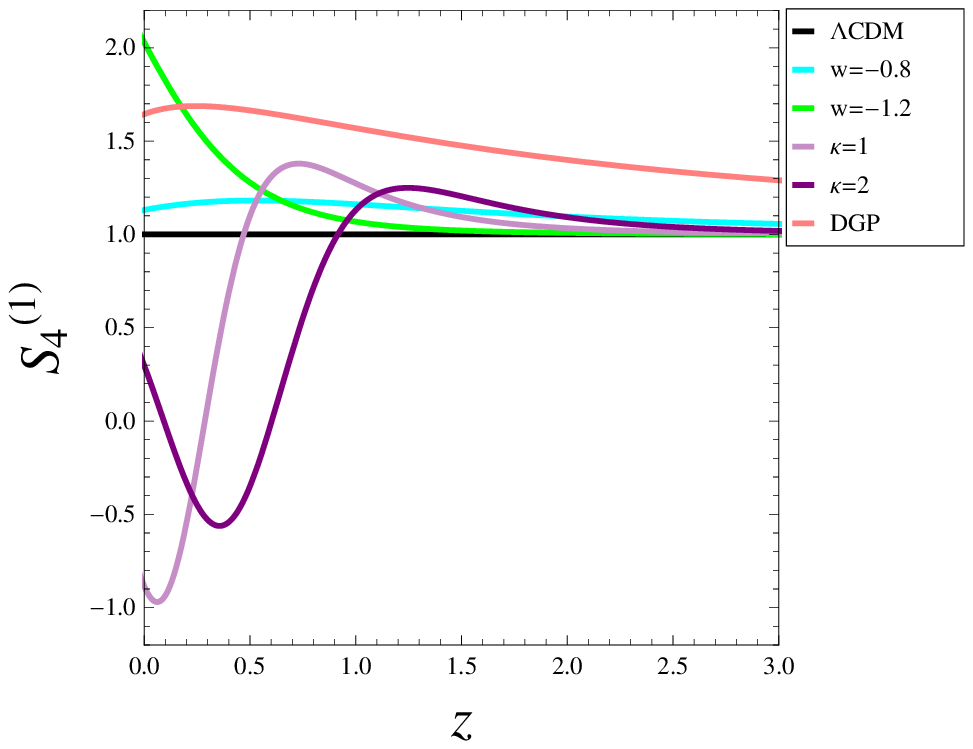}
\end{array}$
\end{center}
\caption{\small 
The left (right) panel shows the Statefinder $S_4^{(1)}$
plotted
against $\om$ (left panel) and $z$ (right panel). 
The dark energy models are: DE with $w = -0.8$ (blue), phantom with
$w = -1.2$ (green),
Chaplygin gas (purple), DGP (red).
The horizontal black line shows $\Lambda$CDM.
The vertical band centered at $\omm = 0.3$ in the left panel roughly
corresponds to the present
epoch. $\omm = 0.3$ is assumed for DE models in the right panel.
} \label{fig3}
\end{figure*}

\begin{figure*}[!t]
\centering
\begin{center}
\vspace{-0.05in}
\centerline{\mbox{\hspace{0.in} \hspace{2.0in}  \hspace{2.0in} }}
$\begin{array}{@{\hspace{-0.3in}}c@{\hspace{0.3in}}c@{\hspace{0.3in}}c}
\multicolumn{1}{l}{\mbox{}} &
\multicolumn{1}{l}{\mbox{}} \\ [-0.5cm]
\includegraphics[scale=1.0, angle=0]{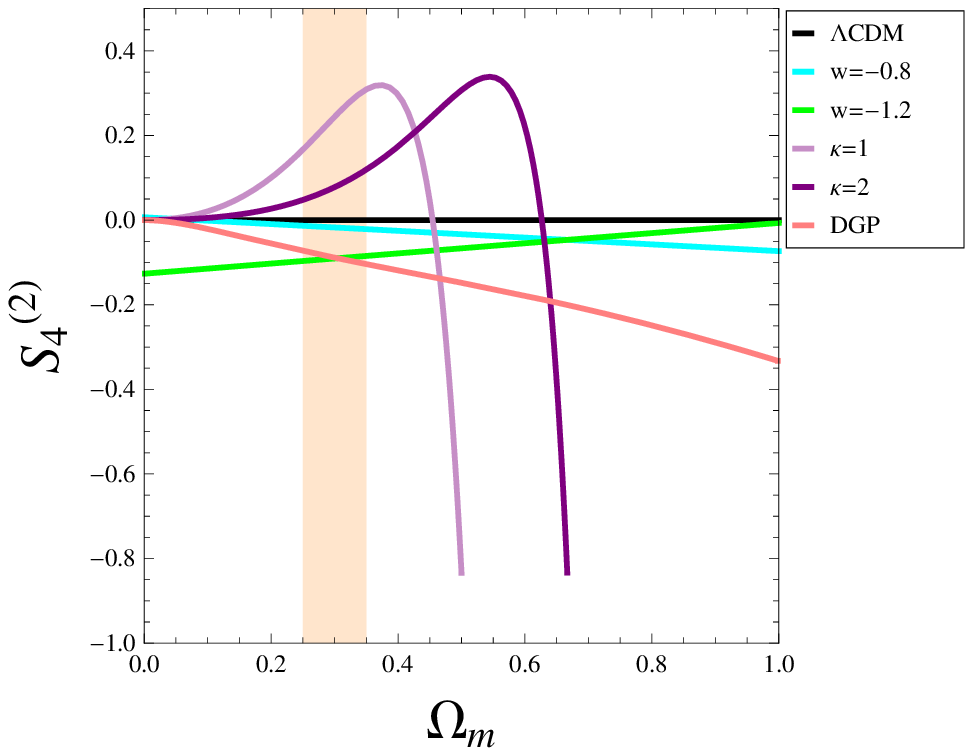}
\includegraphics[scale=1.0, angle=0]{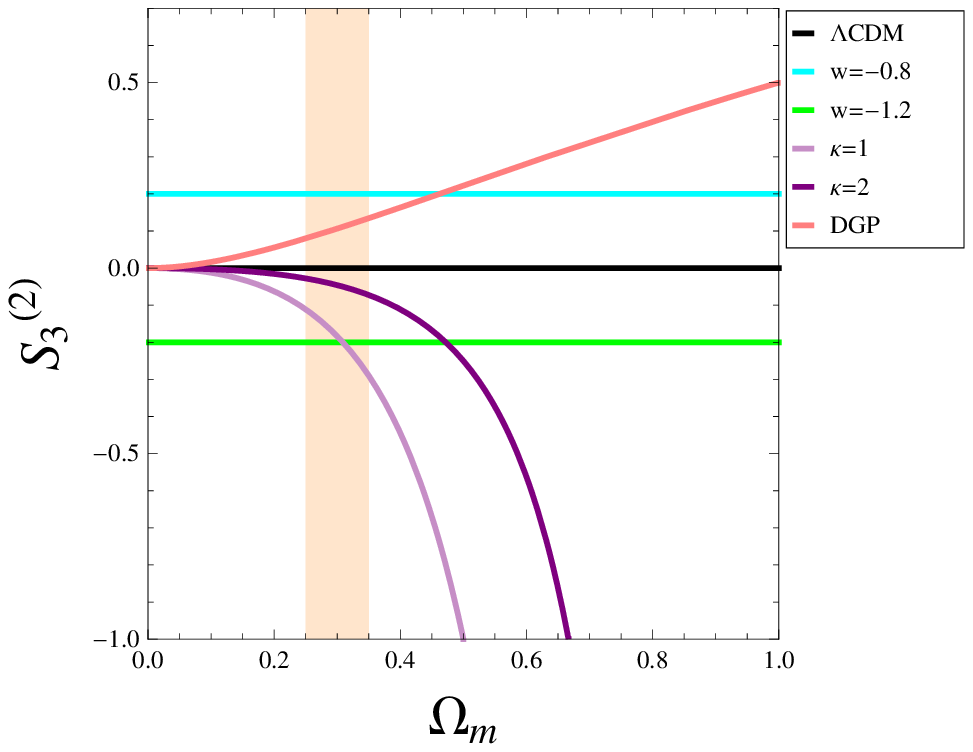}
\end{array}$
\end{center}
\caption{\small
The left (right) panel shows the Statefinder $S_4^{(2)}$ ($S_3^{(2)}$)
plotted
against $\om$ (left panel). 
The vertical band centered at $\omm = 0.3$ in the left panel 
corresponds to the present
epoch.
Comparing figure \ref{fig4} with figure \ref{fig3}, we find that 
$S_4^{(2)}$ does not appear to perform as well as $S_4^{(1)}$,
or even $S_3^{(2)}$, in distinguishing
between the DE
models considered in this paper.
} \label{fig4}
\end{figure*}

\begin{figure}[!t]
\begin{center}
\vspace{-0.1cm}
\psfig{figure=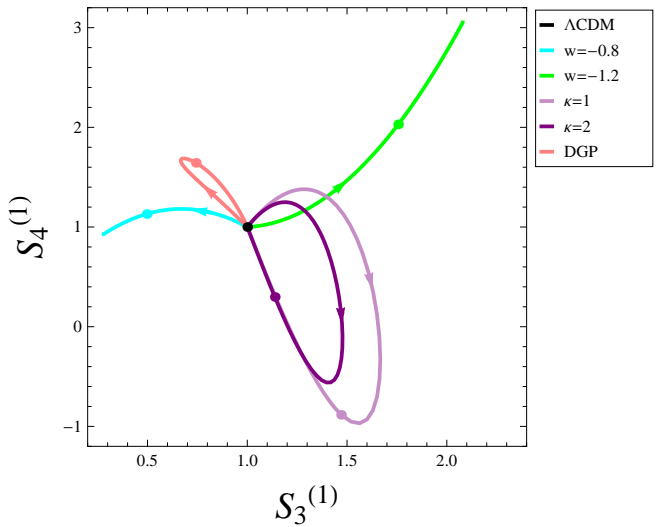,width=0.50\textwidth,angle=0}
\vspace{-0.6cm}
\end{center}
\caption{\small The statefinders
$S_4^{(1)}$ and $S_3^{(1)} \equiv S_3$ are shown for the DE models discussed in the previous
figures.
The fixed point at $\lbrace 1,1\rbrace$ is $\Lambda$CDM. 
The arrows show time evolution and the present epoch in the different models
 is shown as a dot.
$\omm = 0.3$ is assumed.
}
\label{fig5}
\end{figure}

As we demonstrate in figures \ref{fig2}, \ref{fig3}, \ref{fig4}, \ref{fig5},
the Statefinder hierarchy $\lbrace S_n^{(1)},S_n^{(2)}\rbrace$
provide us with an excellent means
of distinguishing dynamical DE models from $\Lambda$CDM.
Our discussion focuses on
the following models:


\begin{enumerate}

\item Dark energy with a constant equation of state
\begin{equation*}
\frac{H(z)}{H_0} = \left\lbrack \omm(1+z)^3 +
\Omega_{\rm DE}(1+z)^{3(1+w)}\right\rbrack^{1/2}~,
\end{equation*}
where concordance cosmology ($\Lambda$CDM) corresponds to $w=-1$.
For simplicity we neglect the presence of spatial curvature 
and radiation.

\item The Chaplygin gas \cite{chap1} has the interesting EOS 
$p_c = - A/\rho_c$ while its density evolves as
$\rho_c=\sqrt{ A+B (1+z)^6}$.
The expansion rate of a universe containing the Chaplygin gas and pressureless
matter is given by
\begin{equation*}
\frac{H(z)}{H_0} = \left\lbrack \omm(1+z)^3 + \frac{\omm}{\kappa}\sqrt{\frac{A}{B} +
    (1+z)^6}\right\rbrack^{1/2}\,\,,
\end{equation*}
where $\kappa$ defines the ratio between the density in 
cold dark matter and the Chaplygin gas
at the commencement of the matter-dominated stage of expansion and
\begin{equation*}
A = B \left\lbrace \kappa^2 \left( \frac{1-\omm}{\omm} \right)^2 - 1
\right\rbrace \,\,.
\end{equation*}

\item The Braneworld model suggested by
Dvali, Gabadadze and Porrati \citep{dgp}
\begin{equation*}
\frac{H(z)}{H_0} = \left[ \left(\frac{1-\omm}{2}\right)+\sqrt{\omm (1+z)^3+
\left(\frac{1-\omm}{2}\right)^2} \right]\,\,.
\end{equation*}

\end{enumerate}

\section{Growth rate of perturbations}

\begin{figure}[ht!]
\begin{center}
\vspace{-0.1cm}
\psfig{figure=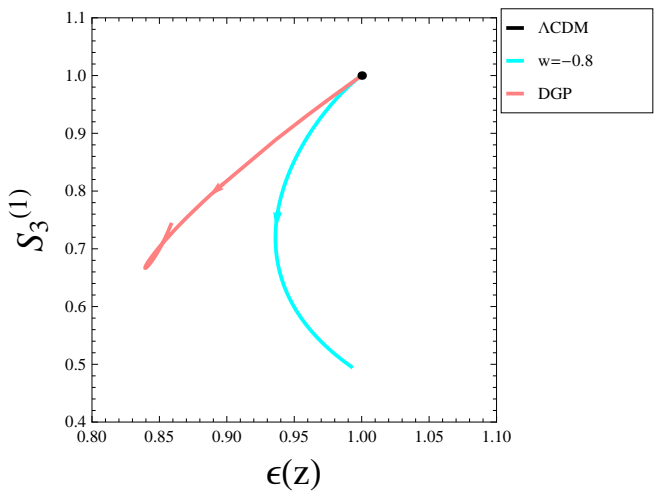,width=0.50\textwidth,angle=0}
\vspace{-0.6cm}
\end{center}
\caption{\small 
The {\it composite null diagnostic} $\lbrace S_3^{(1)}, \epsilon\rbrace$
is plotted for three DE models.
Arrows show time evolution which proceeds from $z=100$ to $z=0$.
}
\label{growth}
\end{figure}

The Statefinders can be usefully supplemented by the
{\it fractional growth parameter} $\epsilon (z)$ \cite{ff3}
\begin{eqnarray*}
\epsilon (z):=\frac{f(z)}{f_{\Lambda CDM}(z)}.
\end{eqnarray*}
where
$f(z) = d\log{\delta}/d\log{z}$ describes the growth rate of linearized
density perturbations 
\cite{growth}
\ber\label{eq:f_w}
f(z) &\simeq& \Omega_m(z)^\gamma\\
\gamma(z) &=& \frac{3}{5-\frac{w}{1-w}}+\frac{3}{125}\frac{(1-w)
(1-\frac{3}{2}w)}{(1-\frac{6}{5}w)^3}\left(1-\Omega_m(z)\right) \nonumber\\
&+&{\cal O}\lbrack\left(1-\Omega_m(z)\right)\rbrack^2 \,\,.
\eer
The above approximation works reasonably well for physical DE models in which
$w$ is either a constant, or varies slowly with time. For instance
$\gamma \simeq 0.55$ for $\Lambda$CDM \cite{growth,ff}.
It may be noted that in physical DE models such as Quintessence, 
the development of perturbations
can be reconstructed from a knowledge of the expansion history \cite{alam09}.
This is not the case in modified gravity theories in which 
perturbation growth 
contains information which is complemenatary to that contained in the
expansion history. (Eqn. (\ref{eq:f_w}) is still
a valid approximation for DGP cosmology but with
$\gamma \simeq 2/3$ 
\cite{f_dgp}.)

For this reason, the 
{\it fractional growth parameter} $\epsilon (z)$, 
 can be used in conjunction with the
Statefinders to define a 
{\it composite null diagnostic}: $CND \equiv 
\lbrace S_n, \epsilon\rbrace$, where $\lbrace S_n, \epsilon\rbrace
 = \lbrace 1, 1 \rbrace$
for \lcdm. 
Figure \ref{growth} demonstrates how DE models 
are distinguished by means of CND.
(One can also use the $Om$ diagnostic in place of the Statefinder in CND.)

\section{Conclusions}
\label{concl}

In this paper we have shown that 
 a simple series of relationships links
the Statefinder hierarchy in concordance cosmology with the deceleration/density
 parameters. 
These relationships can be used to define {\em null tests}
for the cosmological constant. Including information pertaining
to the growth rate of perturbations increases the effectiveness of this
hierarchy of null diagnostics.
Our results demonstrate that lower order members of the
Statefinder hierarchy already differentiate
quite well between concordance cosmology on the one hand, and
Braneworld models and the Chaplygin gas, on the other.
Since, for $n\geq 3$, the $n^{\rm th}$ Statefinder $S_n$ contains
terms proportional to $w^{(n-2)}/H^{n-2}$, higher members of the 
hierarchy will contain progressively greater information about the
evolution of the equation of state of DE.
From the observational perspective, however, one might note that a
determination of the
$S_n$ statefinders involves prior knowledge of the $(n-1)^{\rm th}$
 derivative of $H(z)$.
Thus only lower order Statefinders, $S_n$, $n \leq 4$, together with the
$Om$ diagnostic, may prove compatible with the quality of observational
data expected in the near future.


\section*{Acknowledgments}
We acknowledge useful discussions with Ujjaini Alam and Alexei Starobinsky.

\end{document}